\documentstyle[aps,psfig,twocolumn]{revtex}
\begin{document}
\draft
\bibliographystyle{unsrt}
\title{Shadowing in photo-absorption : role of in-medium hadrons}
\author{Jan-e Alam$^{a}$, Sanjay K. Ghosh$^{a,\ast}$,
Pradip Roy$^{b}$, Sourav Sarkar$^{a}$}   
\address {$^a$ Variable Energy Cyclotron Centre, 1/AF, Bidhannagar,
Calcutta 700 064, INDIA }
\address {$^b$ Saha Institute of Nuclear Physics, 1/AF, Bidhannagar,
Calcutta 700 064, INDIA }
\maketitle
\begin{abstract}
We study the effects of in-medium hadronic properties 
on shadowing in photon-nucleus interactions in Glauber model 
as well as in the multiple scattering approach. 
A reasonable 
agreement with the experimental data is obtained in a scenario 
of downward spectral shift of  the hadrons.  Shadowing is found 
to be insensitive to  the broadening of the spectral functions.  
An impact parameter dependent analysis of shadowing might shed more 
light on the role of in-medium properties of hadrons.
\end{abstract}

\pacs{PACS Nos. : 25.20.Dc,25.40.Ep,24.10.Ht,12.40.Vv}

Theoretical studies based on various models of hadronic interaction
predict a reduction in the mass of hadrons at or above nuclear matter
density~\cite{embr,shl}.
The observation of enhanced dilepton
production in the low invariant mass domain in the heavy ion 
collision experiments~\cite{ceres} does seem to indicate a non-trivial
modification of the properties of light vector mesons, particularly the
$\rho$ meson, in hot and/dense medium~\cite{review}.
However, the complicated dynamics both in the initial as well as final 
states in these experiments inhibits a firm conclusion about nuclear
medium effects at present. 
	
On the other hand, the experiments like photo-absorption on nuclei
provide much cleaner systems for the study of in-medium properties of
mesons \cite{tagx}. With the availability of better photon beams, there 
has been a renewed interest in the photo-absorption processes. 

The phenomena of shadowing plays an important role in the photo-nuclear 
reactions. The photo-nuclear data at lower energies, for different
nuclei, seem to indicate an early onset of shadowing 
\cite{bianchi1,bianchi2}. In ref.\cite{bianchi3} 
this feature has been interpreted 
as a signature for a lighter $\rho$- meson in the medium,
where the shadowing effect was 
evaluated within a Glauber- Gribov multiple scattering theory 
\cite{gribov,weis,glauber1} and generalized vector meson dominance (VMD). 
In contrast the authors in ref.\cite{mosel} 
have claimed that the early onset of shadowing can be understood within
simple Glauber theory \cite{glauber1,glauber2,yennie1,yennie2} if one
takes the negative real part of the $\rho$N scattering amplitude into 
account which corresponds to a higher effective in-medium $\rho$ meson 
mass. In a subsequent paper \cite{mosel1}, the authors 
have concluded that the enhancement of shadowing at low energies occurs 
due to lighter $\rho$ mesons as well as intermediate $\pi^0$ produced in 
non-forward scattering.

In the backdrop of these different inferences, we have made an attempt to
understand the role of in-medium properties of hadrons in the
phenomenon of shadowing in photo-nuclear reactions. 
The shadowing in photon-nucleus reactions can be 
written as,
\begin{eqnarray}
\frac{A_{eff}}{A}=\frac{\sigma_{\gamma\,A}}{A\,\sigma_{\gamma\,N}}
 &=& 1 + \frac{\delta\sigma_{\gamma\,A}}{A\,\sigma_{\gamma\,N}}
\label{cross1}
\end{eqnarray}
where $\sigma_{\gamma A} = A \sigma_{\gamma N} + \delta \sigma_{\gamma A}$
consists of the incoherent scattering of the photon from individual
nucleons and a correction due to the coherent interaction with
several nucleons.The later
has been evaluated using both multiple scattering approach
~\cite{weis,mosel1,piller}
as well as  Glauber's formula along with VMD~\cite{sakurai}.
Resonance contribution to $\gamma-A$ cross-section 
for photon energy $\lesssim 1.2$ GeV has been estimated
by using the prescription given in~\cite{alberico}.

Let us first consider the in-medium effects on the kinematics of
$\gamma-A$ collisions. For a photon incident on a nucleus with
energy $E_{\gamma L}$,
the energy in the rest frame of the nucleon is
given by
\begin{eqnarray}
E_{\gamma} = \gamma_{F} E_{\gamma L} (1-\beta_{F} \cos\theta_{L}),
\label{egamma}
\end{eqnarray}
where $\beta_{F}=p_{F}/E_{F}$
and $\theta_{L}$ is the angle between the incident photon and the Fermi 
momentum of the nucleon which now depends on the space co-ordinate through the
density $n(r)$. The square of the centre of mass  energy $s$, of
the $\gamma-N$ system can then be written as,
\begin{eqnarray}
s &=& (p_\gamma + p_F)^2 \nonumber \\
&=& {m_{N}^{*}}^2 + 2 \gamma_{F} m_{N}^{*} E_{\gamma L} (1-\beta_{F}\cos\theta_{L}),
\label{ss}
\end{eqnarray}
where $m_N^*$ is the effective nucleon mass inside the nucleus. 
The modification of vector meson masses in nuclear environment
has been studied in different models \cite{embr,shl,ghosh,saito}.
Here, we have used two different models namely 
universal scaling scenario (USS)~\cite{embr} and Quantum Hadrodynamical
model (QHD)~\cite{vol16}. In USS,
the effective hadronic masses ($m_{H}^*$) vary 
with nuclear density as
\begin{equation}
\frac{m_H^*}{m_H}=1-0.2x,
\end{equation}
where $x=n(r)/n_0(r)$, $n_0(r)$ being the normal nuclear matter density.
In QHD the effective masses of nucleons and vector
mesons are calculated using standard techniques of thermal field theory
~\cite{ptp,ann} and is parametrized as :
\begin{eqnarray}
\frac{m_H^*}{m_H}=1+\sum_{j=1}\,a_j\,x^j\,.
\end{eqnarray}
For nucleons  $a_1=-0.351277$ and $a_2=0.0766239$;
in case of $\rho$, $a_1=-1.30966$, $a_2=1.78784$, $a_3=-1.17524$ and
$a_4=0.294456$ and finally for $\omega$, $a_1=-0.470454$, $a_2=0.313825$
and $a_3=-0.0731274$. No medium effect on the $\phi$ meson is considered
here as it is expected to be small~\cite{kuwabara}.

At lower energies $\gamma -A$ interaction is known to be
dominated by resonance production~\cite{alberico,kondratyuk}. 
Beyond this region we have used VMD for a better description.
The vector meson produced
inside the nucleus will have an effective mass depending on the
density of the nuclear medium as seen by the meson. 
This change in mass would affect the coherence length 
$\lambda$ which corresponds to the time scale of 
the fluctuation between the bare photon and the hadronic component of the
physical photon. For small $\lambda$, the hadron mediated interaction
may become indistinguishable from the bare photon interaction and 
there will not be any shadowing. In the present case, $\lambda$ becomes 
a function of the radial distance inside the nucleus. For a vector meson 
with effective mass $m_{V}^{*}$, one gets,
\begin{eqnarray}
\lambda = \frac{1}{E_{\gamma} - \sqrt{E_{\gamma}^{2} - {m_{V}^{*}}^2 }}
\,\sim\,\frac{2E_\gamma}{m_V^{*2}},
\label{coherence}
\end{eqnarray}
where $E_{\gamma}$ itself depends on the position of the struck nucleon
through eq.~(\ref{egamma}). Multiple scattering with the nucleons in the
nucleus results in a modified mass of this meson.
The correction to the nuclear photo-absorption cross section due to 
multiple scattering can be written in terms
of $n$-fold multiple scattering amplitude ${\cal{A}}^{(n)}$ as 
\cite{piller},
\begin{equation}
\delta\sigma_{\gamma A}=\frac{1}{2m_{N}k} Im \sum^{A}_{n=2}\,{\cal{A}}^{(n)}
\label{multi1}
\end{equation}
$k$ is the wave vector of the photon and the
$n=1$ term corresponds to the incoherent part. 
In the present work we show results upto double scattering only
(see ref.~\cite{weis,mosel1} for details).

In the high energy limit, under the eikonal approximation, the summation 
of the multiple scattering series goes over to the Glauber's formula if 
one neglects the width of the vector mesons. 
The photon entering the nucleus at an impact parameter $b$ produces a
vector meson at position $z_{1}$.   
Inside the nucleus, the coherence length $\lambda$, in general, would
be different at $z_1$ ($\lambda_1$) and $z_2$ ($\lambda_2$) as the 
different densities will yield different masses. 
The expression for the shadowing part of the cross section 
is then given by~\cite{yennie1},
\begin{eqnarray}
&&\delta \sigma_{VA}=\frac{g_V^2}{4\pi\alpha}\delta\sigma_{\gamma\,A}=\nonumber\\
& & \frac{1}{2k k_{V}}\int{d^2b} \int dz_{1} \int dz_{2}
\exp \left[- \frac{1}{2} \int \sigma_{VN}(z') n(b,z') dz' \right] \nonumber \\
& \times &
k_{V}(z_{1}) \sigma_{VN}(z_{1}) k_{V}(z_{2}) \sigma_{VN}(z_{2}) n^{(2)}(b,z_{1},z_{2})
\nonumber \\
& \times & \left[ \left( \alpha_{V}(z_{1}) \alpha_{V}(z_{2}) - 1 \right)~~ \cos\left(
(\frac{z_{1}}{\lambda_{1}} -
\frac{z_{2}}{\lambda_{2}}) \right. \right. \nonumber \\
&+& \left. \frac{1}{2} \int_{z_1}^{z_2} \alpha_{V}(z') \sigma_{VN}(z') n(b,z') dz' \right)
- \left( \alpha_{V}(z_{1}) + \alpha_{V}(z_{2}) \right)  \nonumber \\
& \times & \left. \sin\left( (\frac{z_{1}}{\lambda_{1}} - 
\frac{z_{2}}{\lambda_{2}}) + \frac{1}{2} \int_{z_1}^{z_2} \alpha_{V}(z') \sigma_{VN}(z')
n(b,z') dz'\right) \right],
\label{deltasig}
\end{eqnarray}
where $\alpha_{V}={Re f_{VV}}/{Im f_{VV}}$ is the ratio of the real and
imaginary part of the $VN$ forward scattering amplitude~\cite{yennie2}. 
$\sigma_{VN}$ 
is the $V-N$ scattering cross section~\cite{yennie2} and $k_{V}$
is the wave vector of the vector meson.
The attenuation of the vector meson amplitude is described by  the 
exponential factor.
We have included 2-body correlation in the 
two-particle density as~\cite{mosel},
$n^{(2)}(b,z_{1},z_{2}) = n(b,z_{1}) n(b,z_{2})
[1-j_{0}(q_{c} |z_{1}-z_{2}|)]$ \\
where $q_{c}=780$ MeV and $j_{0}$ is the spherical Bessel function. 

The authors of ref.~\cite{mosel} have indicated that the scattering 
of the vector meson with the nucleons in the nucleus leads to 
a change in its mass ($\Delta\,m\,\sim\,-2\pi\,n(r)\,Re{f}/m$)~\cite{eletsky}
and concluded that using an external mass would mean an overcounting of the
medium effects. This observation may not be valid entirely. 
Let us consider the QHD model to discuss this point.
It is well known that the small increase in the vector meson mass due to 
its interaction with the Fermi sea is overwhelmed by the large decrease 
due to Dirac sea interaction. This statement, though model dependent, 
does point out that the vacuum fluctuation (VF) which in free space 
renormalizes the particle to its physical mass, may have a different 
role in the medium. Presently, this phenomena can not be described from 
first principles. The increase
in $\rho$ mass due to negative real part of the $\rho -N$ scattering 
amplitude ($\sim$ 10 - 100 MeV~\cite{armstrong,kondra} depending on the 
parameterization) may not be effective enough due to the larger drop from
VF corrections and the net decrease might show up in the experimental 
data. So, while considering the vector meson in the medium, one 
should consider both the effects.  Furthermore, in QHD, a drop in 
nucleon mass causes a larger drop in $\rho$ mass. Hence, for the present 
study it is also necessary to consider the effective nucleon mass inside 
the nucleus which was ignored in the previous studies.
We should mention here that any increase in mass will
lead to a reduction in the shadowing because of the decrease
in $\lambda$. Moreover, the experimental data from
other sources, {\it e.g.} heavy ion collisions~\cite{ceres} 
and proton-nucleus collisions~\cite{kek} seem to indicate a softening
of the vector meson spectral function. 

Before presenting the theoretical results we discuss the available
experimental data. To get the experimental numbers for $A_{eff}$, we have used
$\sigma_{\gamma A}$ from ref~.\cite{bianchi2} and $\gamma$-proton cross
section from refs~.\cite{armstrong,pdg}. The $\gamma$-neutron
cross section is obtained as
$\sigma_{\gamma n} = \sigma_{\gamma d} - \sigma_{\gamma p} + \sigma_{\gamma G}$,
where $\sigma_{\gamma d}$ is taken from ref.\cite{pdg} and $\sigma_{\gamma G}$ 
is the 
Glauber correction which is known to be small at lower energies \cite{yennie2}. The data for 
$\gamma -p$ and $\gamma -n$ are interpolated for the relevant
energies corresponding to given $\sigma_{\gamma A}$ \cite{bianchi2}.
The average photon-nucleon cross section for a nucleus with mass number $A$ 
is given as 
\begin{eqnarray}
\sigma_{\gamma N} = \frac{Z\sigma_{\gamma p} + (A-Z)\sigma_{\gamma n}}{A}
\label{gamn}
\end{eqnarray} 
from which the experimental numbers are obtained as,
\begin{eqnarray}
\frac{A_{eff}}{A} = \frac{\sigma_{\gamma A}}{A \sigma_{\gamma N}}.
\end{eqnarray}

We discuss the results now. Depending on the size of the nucleus
we have used two different density distributions; for $A<16$ 
the shell model density profile of Ref.~\cite{shaw}  
and for heavier nuclei ($A>16$) the density profile from Ref.~\cite{eskola}
has been used.
According to eq.~(\ref{egamma}), $E_\gamma$ is a function
of angle, $\theta_L$ for non-zero $p_F$.
The results which are presented below have been averaged 
over all the angles.  We find that the effect of Fermi momentum 
in the kinematics (through eq.~(\ref{egamma})) is negligibly small.
\begin{figure}
\centerline{\psfig{figure=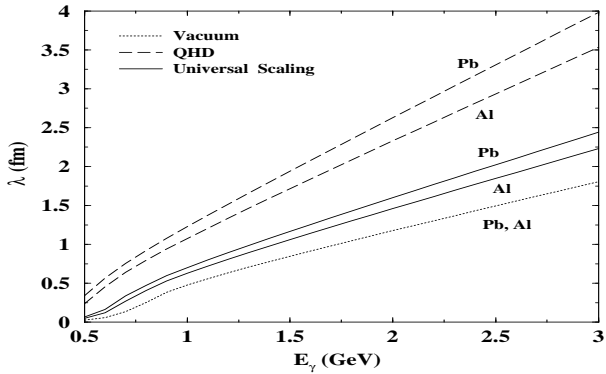,width=8cm,height=5cm}}
\caption{Coherence length for Pb and Al as a function of photon energy 
for various scenarios of effective masses in nuclear medium. 
Solid (dashed) line indicates the coherence length for 
universal scaling (QHD model)and dotted
line shows the results for vacuum mass.
}
\label{colen}
\end{figure}
The variation of $\lambda$ with $E_\gamma$ for $\rho$ mesons
is plotted in  Fig.~\ref{colen} for Pb and Al nuclei.
In order to take both the in-medium mass and width into account
we have folded the coherence length with the spectral function
(as in eq.~(\ref{rhosp}) below). We observe that $\lambda$ is larger 
at lower values of $E_\gamma$ in the QHD scenario than in the
universal scaling approach. Again the heavier nuclei 
seem to be affected earlier. These observations are crucial in the 
understanding of shadowing effects.

Fig. 2 shows the variation of $A_{eff}/A$ with $E_{\gamma}$ for different 
nuclei. The region below 1.2 GeV can be well described by the contribution 
of the baryonic resonances alone. For the region
$1.2 < E_{\gamma} < 3$ GeV we have used VMD with (Glauber) and
without (multiple scattering) eikonal approximation. We find that the
USS gives a better description of the data both in the
multiple scattering approach  and Glauber model than the scenario
with vacuum properties of hadrons.
In QHD the drop of mass
being larger, in general, the data is underestimated. It is also necessary
to point out that though the resonance and VMD regions have been 
differentiated by a vertical line at $E_{\gamma} = 1.2$ GeV, there might as 
well be an overlap of both the pictures around this point. On an average, 
our results
show that the experimental data over the entire range of photon
energy under consideration are reasonably well reproduced by 
the downward shift of the spectral function within the USS.

\begin{figure}
\centerline{\psfig{figure=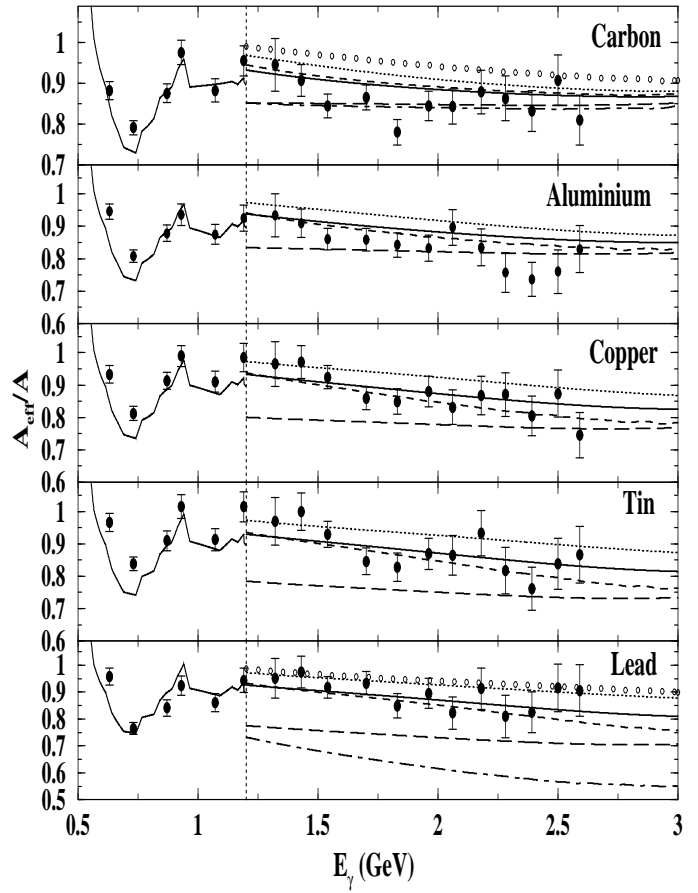,width=9cm,height=12cm}}
\caption{$A_{eff}/A$ for various nuclei as a function of photon
energy. For $E_\gamma\,<\,1.2$ GeV the results
for the baryonic resonance contribution are shown. For photon
energy $\geq 1.2$ GeV we show the results for both multiple
scattering approach and Glauber model. The dotted, long-dashed and
solid lines indicate calculations using Glauber model for vacuum, QHD and
USS respectively. The circles, dot-dashed (shown for C and Pb) and
short-dashed lines correspond to the same in the multiple scattering approach.
}
\label{ctopb}
\end{figure}

\begin{figure}
\centerline{\psfig{figure=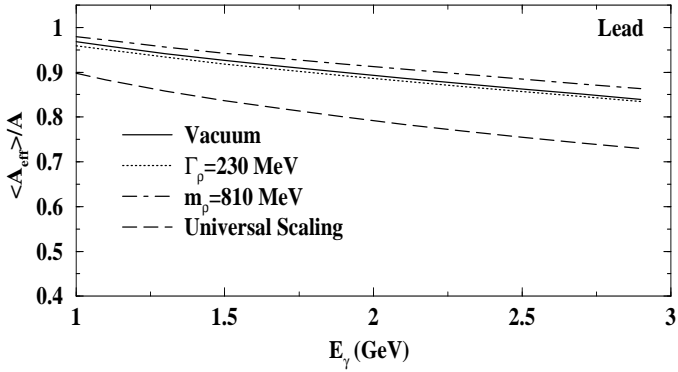,width=9cm,height=5cm}}
\caption{$\langle\,A_{eff}\,\rangle/A$ as a function of photon energy
for lead nucleus.}
\label{spfn}
\end{figure}

As mentioned before the shift in the hadronic spectral
function in the nuclear medium is an unsettled issue.
The experimental data on dilepton production from heavy ion
collisions at CERN super-proton synchrotron energies
can be explained either by shifting the pole mass ($m_V$) 
to a lower value or by increasing the width ($\Gamma_V(M)$) of the
spectral function. 
We would like to demonstrate here how these kind 
of medium effects (pole mass shift or broadening) affect the shadowing. 
For this we have considered the quantity,
\begin{equation}
\langle\,A_{eff}\rangle=\frac{\int\,\rho(M)\,A_{eff}(M)\,dM}{\int\,\rho(M)\,dM}
\label{avaeff}
\end{equation}
where 
\begin{equation}
\rho(M)=\frac{1}{\pi}\,\frac{M\Gamma_V(M)}{(M^2-m_V^{*2})^2\,+\,M^2\Gamma_V^2(M)}\,.
\label{rhosp}
\end{equation}
$A_{eff}(M)$ gets maximum weight 
at the peak of the spectral function, {\it i.e.} from the point $M^2=m_V^{*2}$
and the contribution from either side of this point is approximately 
averaged out. Therefore,
the results become sensitive to the pole mass and is largely 
insensitive 
to the broadening of the spectral function.
In fig.~\ref{spfn} we show the quantity $\langle\,A_{eff}\,\rangle/A$ as a
function of photon energy for $\rho$-meson only. As explained above
the results for vacuum mass 770 MeV and widths 150 MeV (solid line)
and 230 MeV~\cite{kondra} (dotted line) are indistinguishable. However,
a pole mass shift in USS shows a larger shadowing as $\lambda$ 
increases substantially in this case. Moreover, an increase  ($\sim$ 40 MeV 
~\cite{eletsky})
in $\rho$ mass results in the decrease in shadowing (dot-dashed line).
 
\begin{figure}
\centerline{\psfig{figure=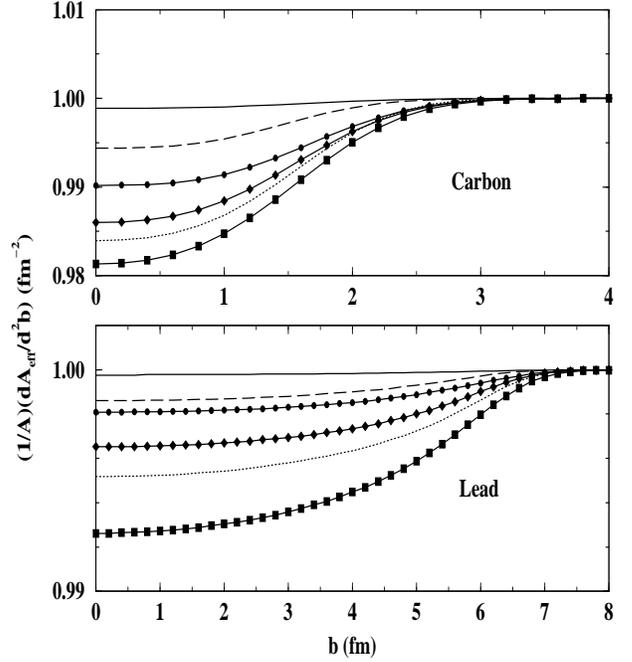,width=8cm,height=9cm}}
\caption{Impact parameter dependence of the shadowing factor.
Solid, dashed and dotted lines are the results for  
vacuum, USS and QHD respectively for $E_\gamma=1$ GeV. 
Filled circle, diamond and square are the corresponding results for 
$E_\gamma=2$ GeV.  
}
\label{impact}
\end{figure}

In all the results shown above, the shadowing is evaluated by 
integrating over all the
values of impact parameter. 
On the other hand, an incident photon  
passing through the nucleus peripherally would see less shadowing 
due to lower densities.
To visualize it we have plotted  
$d(A_{eff}/A)/d^2 b$ with impact parameter in fig.~\ref{impact}. 
While going from C to Pb, we observe that at 
the lower impact parameter shadowing is larger for lighter nuclei. 
This phenomena is a reflection of the nuclear density profile, which
for lighter nuclei is larger in the core region compared to the heavier 
nuclei. It will be interesting to know whether one can define a centrality
parameter for $\gamma - A$ collisions as is usually done for heavy ion 
collisions ( percentage minimum bias etc.).

To conclude, we have studied the effects of in-medium properties of hadrons on
shadowing in photo-absorption processes both in the framework of Glauber
model and multiple scattering approach. 
The general pattern of experimental data seem to
prefer a dropping vector meson mass scenario. 
The universal scaling appears to be closer to the data. 
The shadowing effect is insensitive to the spectral broadening of the
vector meson in the nuclear medium. In contrast to the previous works the 
spatial dependence of the masses of both vector mesons and nucleon 
are considered here. The effect of Fermi motion is found to be small in the
kinematics of the process. 
However, the effect of two-body correlation is important as its absence  
overestimates the data. 
We would also like to comment on QHD. The simple Walecka model,
which we have used here has its own limitations (e.g. large
incompressibility etc.). In this model, the reduction in the 
nucleon and vector meson masses is substantially larger
than other models, which leads to large amount of shadowing.

Finally, we emphasize that 
our understanding of the shadowing phenomena vis-a-vis 
Glauber model may be improved through corrections to the 
approximations inherent in the model as well as by examining 
the model parameters critically.
The leading correction to the Glauber model due to deviation
from eikonal propagation gives rise to a correction $\sim (2P_{cm}^AR)^{-1}$
relative to the Glauber scattering amplitude~\cite{sjw}. Here $P_{cm}^A$ is the
centre of mass momentum of the nucleus and $R$ is its charge
radius. In the present case the correction is rather small
for $E_\gamma>1$ GeV. 
Moreover, a refinement of the Glauber model (multiple
scattering) parameters, {\it e.g.}, vector meson-nucleon scattering
amplitude, two nucleon correlation etc. might give a good agreement with
the data even with the vacuum properties of the hadrons. Hence a better
estimate of these quantities is essential for
a definitive statement
regarding the role of medium effects on  
shadowing in photo-absorption processes.  
Experimental data with better statistics would certainly help 
us to resolve these uncertainties.


\end{document}